\documentclass[aps,prl,twocolumn,showpacs]{revtex4}
\usepackage[dvips]{graphicx}
\usepackage{amsmath}
\usepackage{times}
\usepackage{psfrag}

\begin{document}

\title
{
Interference of atomic levels and
superfluid -- Mott insulator phase transition in a two-component Bose-Einstein
condensate
}

\author
{K.~V.~Krutitsky
and R.~Graham
}

\affiliation{
Fachbereich Physik der Universit\"at Duisburg-Essen, Standort Essen,
Universit\"atsstr. 5, Postfach 10 37 64, 45117 Essen, Germany
}

\date{\today}

\begin{abstract}
The superfluid -- Mott insulator phase transition
in a Bose-Einstein condensate of neutral atoms with doubly degenerate internal
ground states in an optical lattice is theoretically investigated.
The optical lattice is created by two counterpropagating linearly polarized
laser beams with the angle $\theta$ between the polarization vectors (lin-angle-lin
configuration). The phase diagram of the system and the critical values
of the parameters are worked out. It is shown that the sign of the detuning plays
an important role and that there is a strong suppression of the Mott transition
in the case of blue detuning. Varying the laser intensity and/or the angle $\theta$
one can manipulate the Mott-insulator to
superfluid quantum phase transition as well as
prepare the condensate in physically distinguishable ``ferromagnetic" and
``antiferromagnetic" superfluid states.
\end{abstract}


\maketitle

\psfrag{t}[b]{\rotatebox{180}{$2J/U$}}
\psfrag{m}[b]{\rotatebox{180}{${\mu}/{U}$}}
\psfrag{f}[c]{$\theta$}

\psfrag{a}[c]{$(a)$}
\psfrag{b}[c]{$(b)$}
\psfrag{c}[c]{$(c)$}
\psfrag{d}[c]{$(d)$}

The study of quantum phase transitions (QPT) in optical lattices has become
one of several focusses of current interest in the ongoing exploration
of the rich physics of Bose-Einstein condensates (BEC).
In the seminal paper by Jaksch {\it et al.}~\cite{Jaksch} it was predicted that
by increasing the strength of the periodic optical potential one can confine
ultracold atoms at lattice sites, which leads to the superfluid--Mott insulator
phase transition, and this effect has been experimentally observed in a one-component
BEC of rubidium atoms~\cite{Greiner}. Recently coherent transport of multicomponent
BEC in optical lattices has been demonstrated~\cite{MGW}.
In such systems a number of new phenomena associated with QPT are possible.
Theoretical studies show that by using different laser schemes one can create
coexisting superfluid and Mott phases~\cite{Jaksch,CW},
fragmented condensates with topological
excitations~\cite{DZ}, maximally entangled atomic states~\cite{You},
dimer phases~\cite{Yip}, and heteronuclear molecular BECs~\cite{MS}.

One of the most intriguing features of the multicomponent BECs is
the structure of their internal levels. The action of lasers on multi-level atoms
can lead to interference of atomic levels. This kind of interference has been
used in atomic BECs to slow down the light~\cite{Hau}, which allows us to excite
quantum shock waves in a BEC~\cite{DBSH}.
In the present Letter we shall investigate the possibility of using interference
of atomic levels for the manipulation of QPT
in a two-component BEC with spatially periodic coupling of the degenerate internal ground
states. It is shown that one can effectively manipulate QPT
not only by varying the laser intensity, but also the laser polarization,
and one can even change the sign of the tunneling matrix element, which leads to
``ferromagnetic" and ``antiferromagnetic" states in analogy to the
spin ordering in magnetic systems.

We consider a two-component BEC of neutral polarizable atoms of mass $M$,
possessing an excited electronic state characterized by the magnetic quantum number $m = 0$
and two Zeeman-degenerate internal ground states with $m = \pm 1$ ($F_g=F_e=1$),
in a one-dimensional optical lattice.
The latter is assumed to be created by two counterpropagating
linearly polarized
laser waves of equal amplitudes and frequencies
with the wave number $k_L$, and the angle $\theta$ between the polarization vectors
(lin-angle-lin configuration)~\cite{LAL}, and with detuning  $\Delta$ from
the internal atomic transition. In order to avoid decoherence due to
spontaneous emission,
$|\Delta|$ must be much larger than the spontaneous emission rate.
The running laser waves form left- and right- polarized standing waves with
the Rabi frequencies
$
\Omega_\pm
=
\Omega_0
\cos( k_L z \pm \theta/2)
$.
Because of the large detuning the excited state can be adiabatically eliminated.
The resulting effective Hamiltonian  couples the atomic ground states and has
the following form:
\begin{equation}
\label{H}
H
=
\int
\Psi^\dagger\cdot
\left[
-
\frac{\hbar^2}{2M}
\frac{\partial^2}{\partial z^2}
+
\frac{\hbar}{\Delta}
\hat\Omega^2
+
\frac{g}{2}
(\Psi^\dagger\cdot
\Psi)
\right]
\cdot
\Psi
\,dz
\,,
\end{equation}
where $\Psi$ is a two-component spinor-field operator.
$\hat\Omega^2$ determines the periodic potential with the period $\pi/k_L$.
It is a $2 \times 2$ matrix with the components
$\hat\Omega^2_{\alpha\beta}=\Omega_\alpha\Omega_\beta$,
$\alpha,\beta=+,-$.
The parameter $g$ describes the repulsive interaction of the condensate
atoms. In the one-dimensional case it has the form~\cite{1DBEC}
$
g
=
{4 \hbar^2 a}/{M a_\perp^2}
\left(
1-1.4603\,{a}/{a_\perp}
\right)
$,
where $a$ is a symmetric scattering length and
$a_\perp=\sqrt{2\hbar/M\omega_\perp}$ is the size of the ground state
for the harmonic potential with the frequency $\omega_\perp$
confining the BEC in the transverse directions.

\begin{figure}[t]
\centering
\hspace{-4mm}  \includegraphics[width=4.4cm]{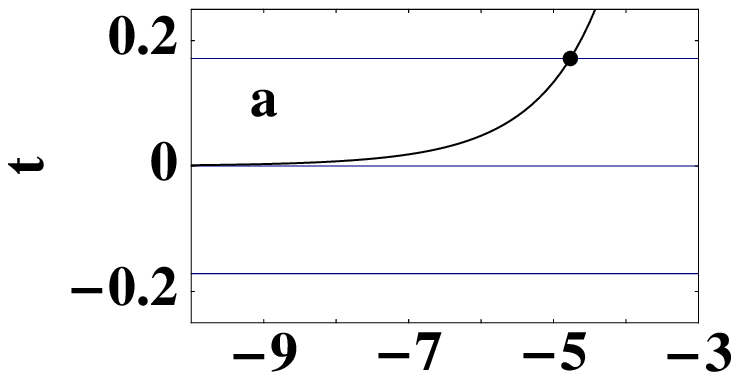}
\hspace{-8mm}  \includegraphics[width=4.4cm]{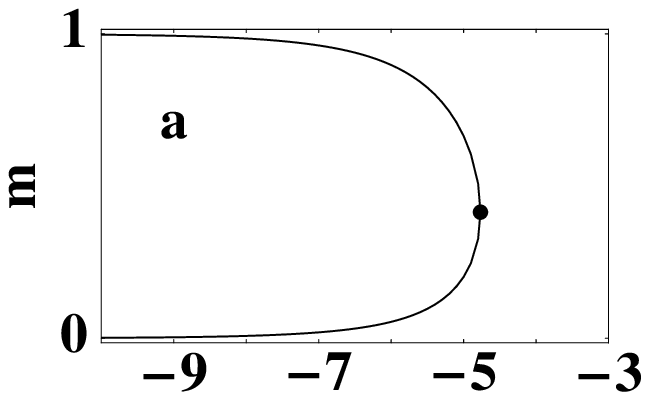}\\
\vspace{-6mm}
\hspace{-4mm}  \includegraphics[width=4.4cm]{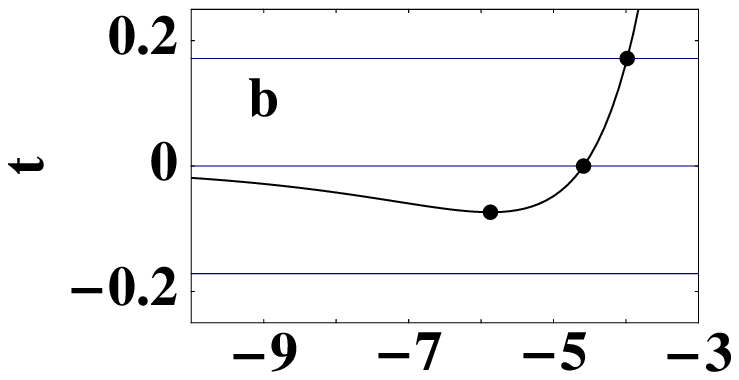}
\hspace{-8mm}  \includegraphics[width=4.4cm]{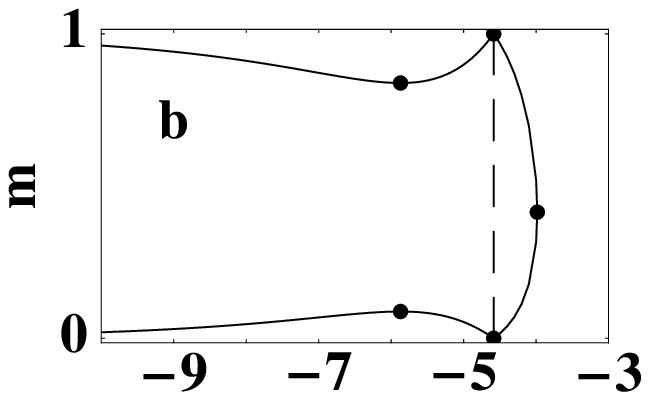}\\
\vspace{-6mm}
\hspace{-4mm}  \includegraphics[width=4.4cm]{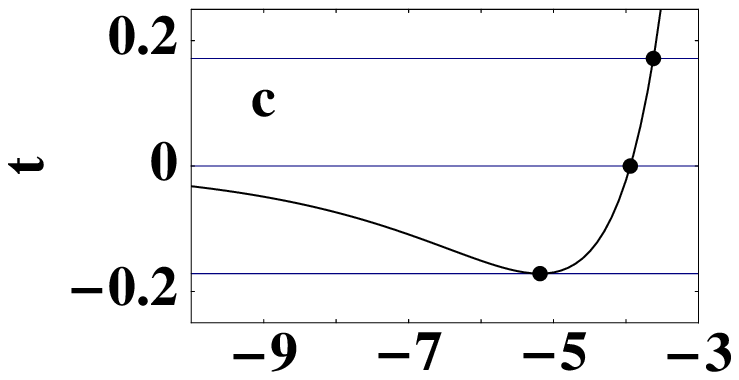}
\hspace{-8mm}  \includegraphics[width=4.4cm]{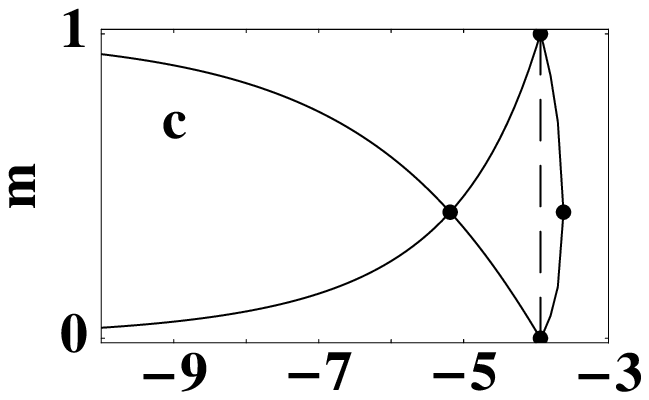}\\
\vspace{-6mm}
\hspace{-4mm}  \includegraphics[width=4.4cm]{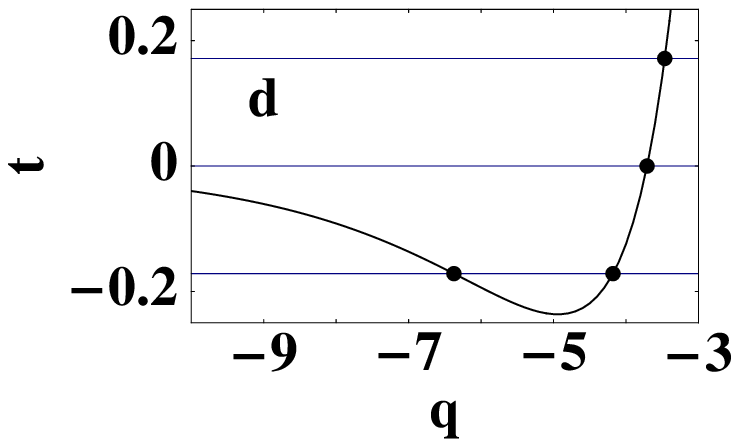}
\hspace{-8mm}  \includegraphics[width=4.4cm]{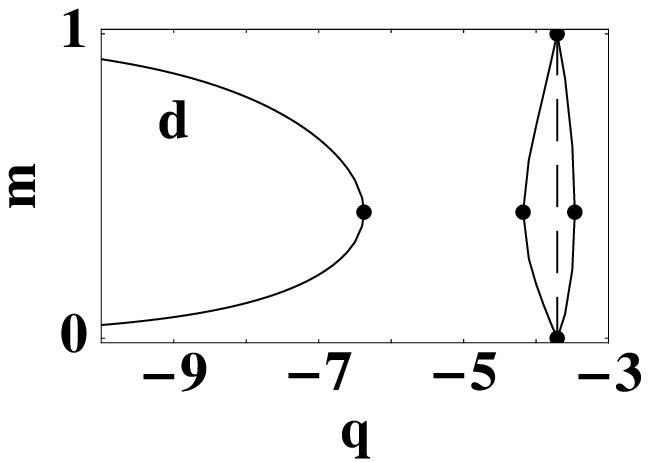}\\
\caption{Typical dependences of $J/U$ on the dimensionless effective Rabi frequency
         $q={\Omega_0^2}/{4\omega_R\Delta}$ for red detuning ($\Delta<0$)
         and corresponding phase diagrams for $n=1$.
	 $\theta=50^\circ$ (a), $64^\circ$ (b), $65.7767^\circ$ (c), $66.5^\circ$ (d).
	 Horizontal straight lines show the critical values
	 $(2J/U)_c = \pm (3-2\sqrt{2})$.
	 Vertical dashed lines on the phase diagrams (right column) indicate
	 the boundary between ferromagnetic ($J>0$) and antiferromagnetic ($J<0$)
	 phase ordering. The regions of the Mott phase are enclosed by the lobes and loops.
        }
\label{phdq}
\end{figure}

\begin{figure}[t]
\centering
\hspace{-4mm}  \includegraphics[width=4.4cm]{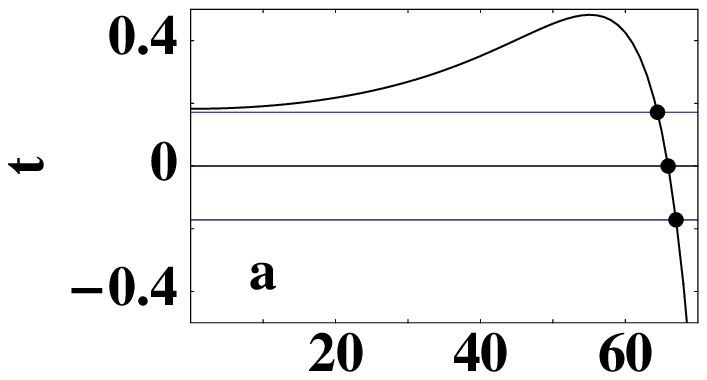}
\hspace{-8mm}  \includegraphics[width=4.4cm]{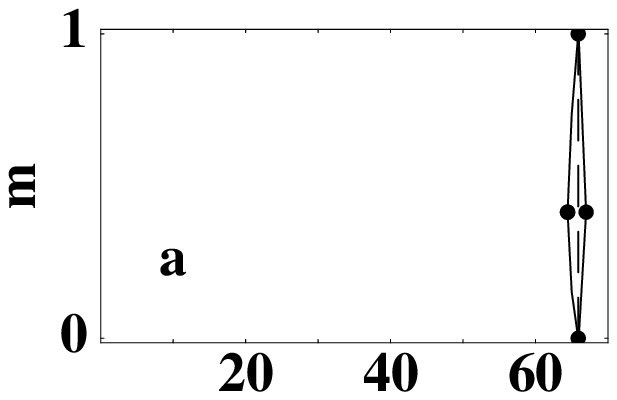}\\
\vspace{-6mm}
\hspace{-4mm}  \includegraphics[width=4.4cm]{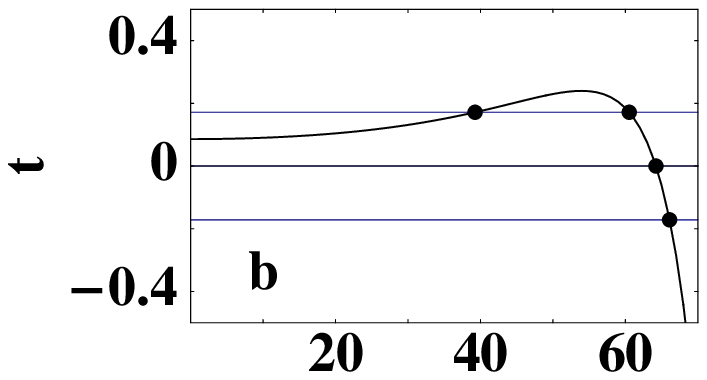}
\hspace{-8mm}  \includegraphics[width=4.4cm]{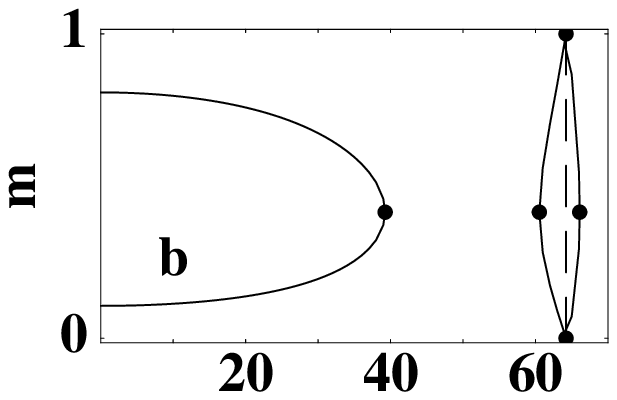}\\
\vspace{-6mm}
\hspace{-4mm}  \includegraphics[width=4.4cm]{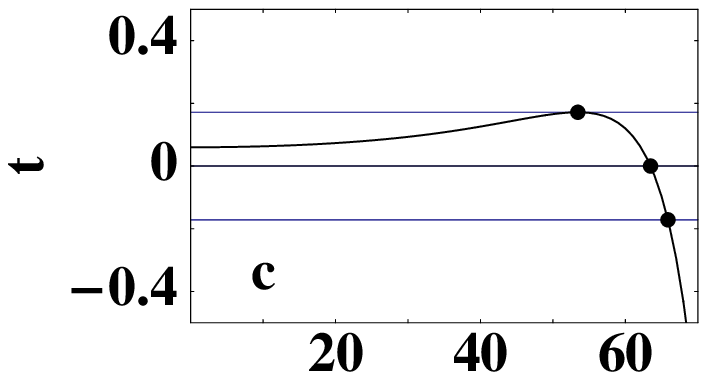}
\hspace{-8mm}  \includegraphics[width=4.4cm]{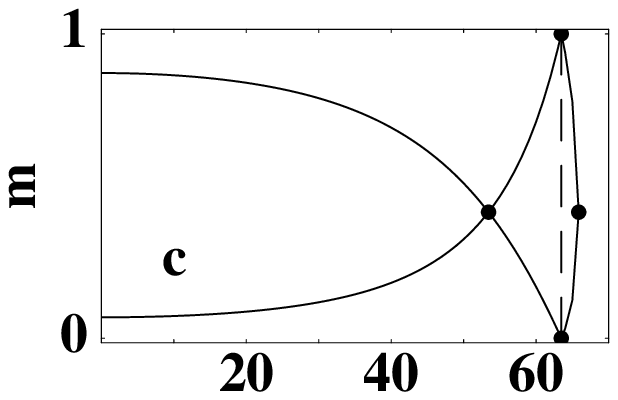}\\
\vspace{-6mm}
\hspace{-4mm}  \includegraphics[width=4.4cm]{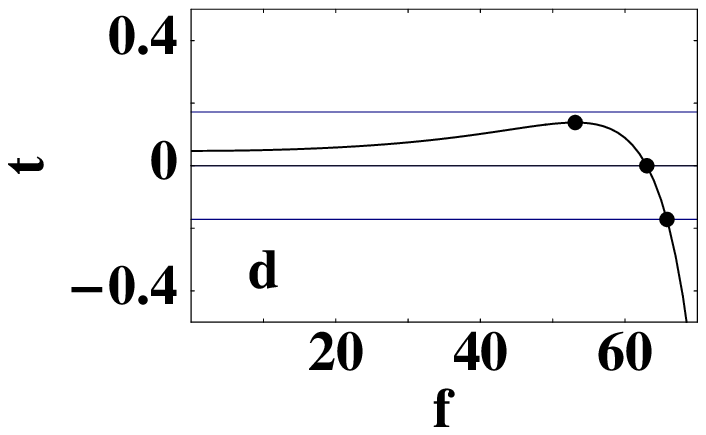}
\hspace{-8mm}  \includegraphics[width=4.4cm]{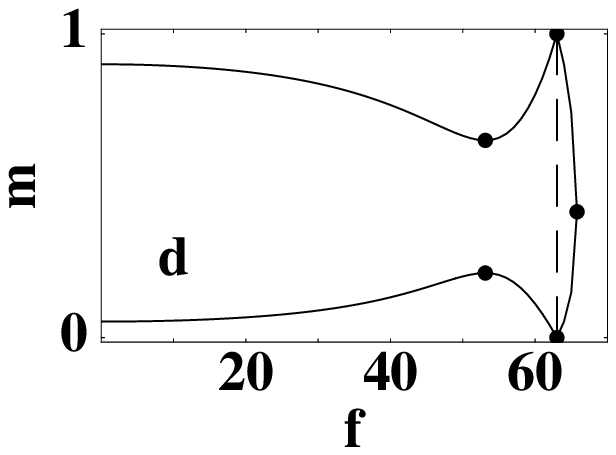}\\
\caption{The same as in Fig.~\ref{phdq} as functions of $\theta$ ($\Delta<0$).
	 $q=-3.9$ (a), $-4.5$ (b), $-4.80064$ (c), $-5$ (d).
        }
\label{phdf}
\end{figure}

\begin{figure}[t]
\centering
\hspace{-4mm}  \includegraphics[width=4.4cm]{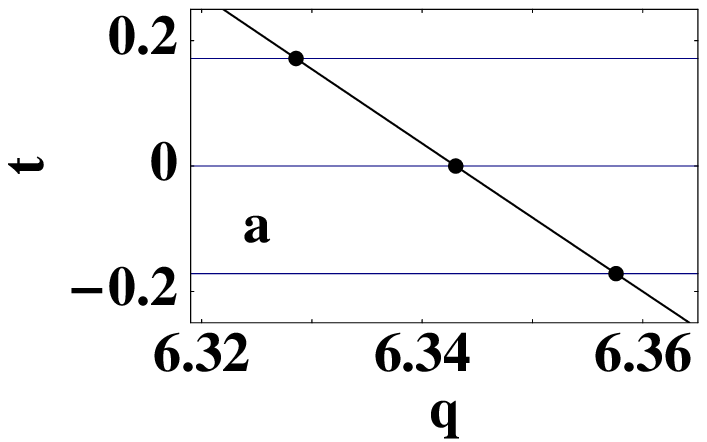}
\hspace{-8mm}  \includegraphics[width=4.4cm]{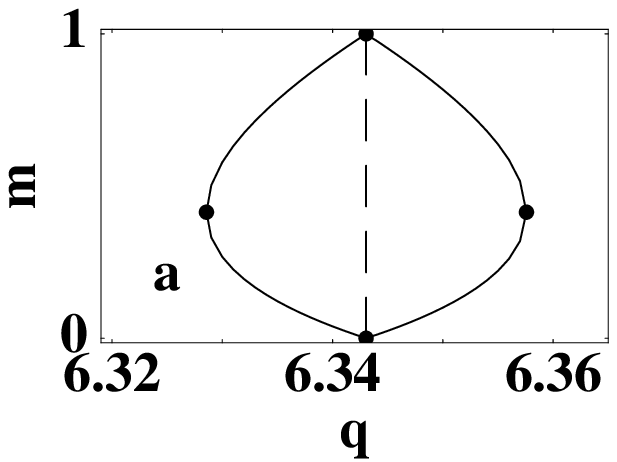}\\
\vspace{-2mm}
\hspace{-4mm}  \includegraphics[width=4.4cm]{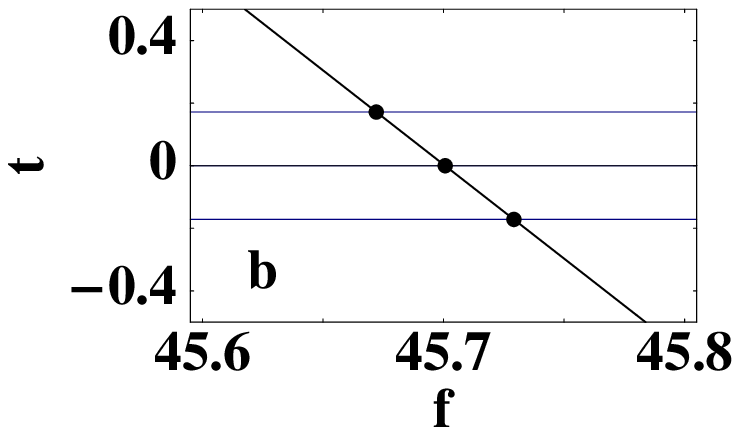}
\hspace{-8mm}  \includegraphics[width=4.4cm]{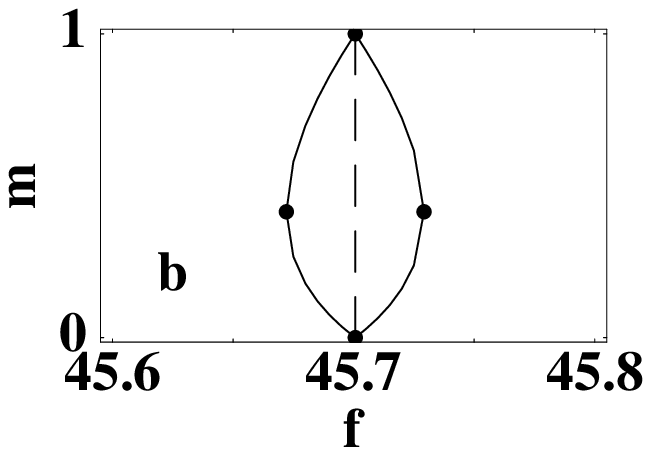}\\
\caption{The same as in Figs.~\ref{phdq} and~\ref{phdf} for blue detuning ($\Delta>0$).
	 $\theta=45^\circ$ (a), $q=6$ (b).
        }
\label{phd-pq}
\end{figure}

The starting point for the investigation of QPT is
the Bose-Hubbard model. We assume that the atoms are prepared in the lowest
Bloch band and the matter-field operator $\Psi$ can be decomposed
in the Wannier basis~\cite{Wannier}
\begin{equation}
\label{Psi}
\Psi(z)=
\sum_i
\exp
\left(
   i \varphi_i
\right)
{\bf W}_i(z) a_i
\,,
\end{equation}
where ${\bf W}_i(z)={\bf W}(z-z_i)$ are two-component Wannier spinors for the
lowest energy band.
They are obtained by the solution of the eigen-value problem for the Hamiltonian
(\ref{H}) in the case $g=0$ and satisfy the orthonormality condition
$\int
{\bf W}_i^\dagger(z)\cdot
{\bf W}_j(z)
\,dz
=
\delta_{ij}
$.
The indeces $i,j$ label the sites of the one-dimensional periodic lattice.
The phases $\varphi_i$ are not yet defined and their proper choice is
discussed in a moment. The $a_i$ and their adjoints are Bose annihilation and
creation operators attached to the lattice sites.
Substituting (\ref{Psi}) into (\ref{H}) and taking into account
only the hopping between the nearest neighbour lattice sites and the atomic
interactions at the same lattice site, we obtain the well-known
Bose-Hubbard Hamiltonian
\begin{eqnarray}
\label{BHH}
H_{BH}
&=&
-J\sum_{<i,j>}
a_i^\dagger
a_j
\exp
\left[
    i
    \left(
        \varphi_j-\varphi_{i}
    \right)
\right]\\
&+&
\frac{U}{2} \sum_i n_i (n_i -1)
-\mu \sum_i n_i
\,,
\nonumber
\end{eqnarray}
where $\mu$ is a chemical potential.
The tunneling matrix element
$
J
=
-\int
{\bf W}_{i+1}^\dagger(z)\cdot
\left(
-
\frac{\hbar^2}{2M}
\frac{\partial^2}{\partial z^2}
+
\frac{\hbar}{\Delta}
\hat\Omega^2
\right)\cdot
{\bf W}_i(z)
\,dz
$
and the atomic interaction parameter
$
U
=
g
\int
{\bf W}_i^\dagger
\cdot
\left(
{\bf W}_i^\dagger
\cdot
{\bf W}_i
\right)
\cdot
{\bf W}_i
\,dz
$
can be simultaneously changed by varying the effective Rabi frequency
and/or the angle $\theta$, but the variation of $J$ is much faster.
Typical dependences of the ratio $J/U$ on the
dimensionless effective Rabi frequency $q=\Omega_0^2/4\omega_R\Delta$,
and the angle $\theta$ are shown in the left column of
Figs.~\ref{phdq},~\ref{phdf}, and~\ref{phd-pq}.
$\omega_R=\hbar k_L^2/2M$ is a one-photon recoil frequency.
The calculations are performed for $^{87}$Rb
($M = 1.45 \times 10^{-25}\ kg$, $a=5.4\ nm$~\cite{Ho}), $\omega_\perp=2\pi\times 200\ Hz$,
$\lambda_L=2\pi/k_L=780$ nm.

In our model we neglect the antisymmetric scattering length, which is responsible
for
(i) the asymmetry of the interatomic collisions for the atoms in the same and in
different ground states
and (ii) inelastic collisions of the atoms in the states $m=-1$ and $m=1$, which
lead to the creation of the atoms in the ground state $m=0$.
This approximation is well justified at least for $^{87}$Rb, because the asymmetry
leads to a correction of the parameter $U$ of the order of only a
few percent in Eq.(\ref{BHH})
and the inelastic collisions become important for the times $T\gtrsim 3/n$ s,
where $n$ is the number of atoms per lattice site.

The phases $\varphi_i$ are determined from the requirement of
minimal energy of the Hamiltonian (\ref{BHH}), which amounts to demanding
that $J\cos(\varphi_i-\varphi_{i+1})$ is maximal. In the familiar case of a
one-component BEC in a periodic lattice,
where the Wannier spinors reduce to functions,
this leads to $\varphi_i=\varphi_j$, because in such a situation $J$ in
Eq.(\ref{BHH})
is real and positive. Therefore the superfluid state of a one-component BEC
in the optical lattice always has a ferromagnetic ordering of the phases
$\varphi_i$, which can then all be set to zero. Interestingly,
in the case of a two-component BEC $J$ can also become negative.
In this case the choice $\varphi_i=\varphi_{i+1}$ does not
provide a minimal energy and $\varphi_{i+1}=\varphi_i\pm\pi$ has to be chosen
instead. This corresponds to an antiferromagnetic ordering of the phases of
neighboring lattice sites. Indeed, as one can see in
Figs.~\ref{phdq},~\ref{phdf}, and~\ref{phd-pq} (left column)
the parameter $J$ takes in general positive as well as negative
values ($U$ is always positive for a BEC with repulsive interaction).
Therefore, ferromagnetic and
antiferromagnetic phase ordering occurs in different parts of the
phase diagram.

Ferromagnetic and antiferromagnetic phase ordering is possible only
for the superfluid phase. These two cases are readily
distinguishable experimentally in the superfluid regime via the
spatial interference pattern generated by the
coherent matter waves which one obtains after removing the lattice potential
\cite{Greiner}:
The interference maxima obtained in the ferromagnetic case turn into
minima in the antiferromagnetic case and vice versa.
In the Mott phase, the numbers of particles occupying each lattice site
are equal and integer, and the phases of the corresponding wave functions
are completely undefined.
Therefore, the choice of the $\varphi_i$ remains arbitrary in this case and has
no observable consequences for the interference pattern.

\begin{figure}[t]
\centering
  \includegraphics[width=4.cm]{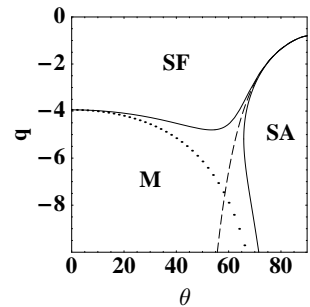}

\caption{Phase diagram in the $(\theta,q)$ plane for $(\mu/U)_c=\sqrt{2}-1$ ($\Delta<0$).
         The boundaries between the superfluid and the Mott ($n=1$) phases are shown
	 by solid (calculation for the two-component BEC) and dotted
	 (approximation, which does not take into account the coupling between
	 the bright and dark states) lines. The dashed line corresponds to $J=0$.
	 It lies always in the Mott phase, separating
	 the regions of the ferromagnetic (SF) and antiferromagnetic (SA)
         superfluid phases.
        }
\label{qf}

\end{figure}

\begin{figure}[t]
\centering
  \includegraphics[width=4.cm]{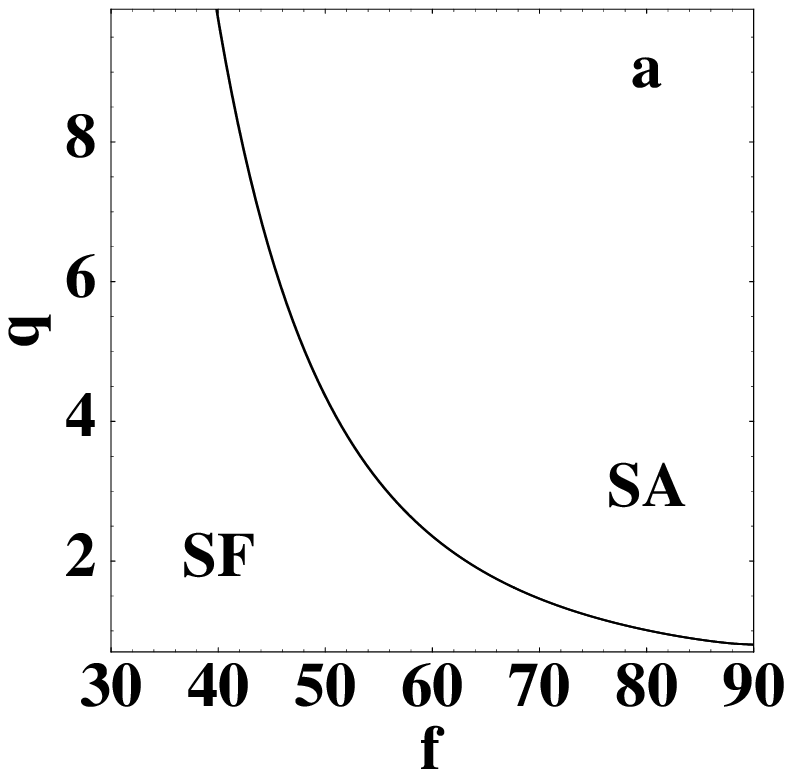}
  \includegraphics[width=4.3cm]{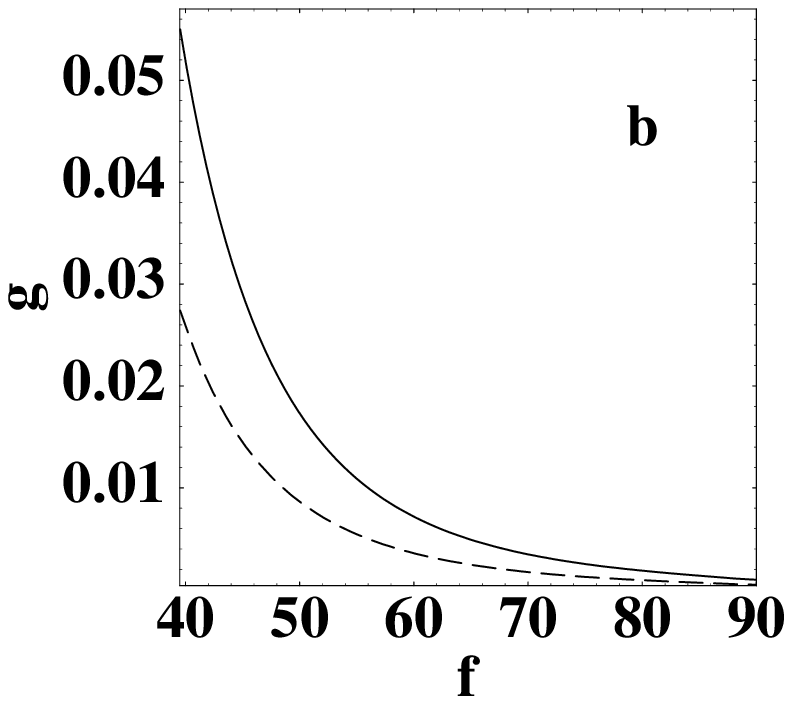}

\caption{(a) Phase diagram in the $(\theta,q)$ plane for $(\mu/U)_c=\sqrt{2}-1$, $\Delta>0$.
         The line $J=0$ as well as the two boundaries separating SF
	 and SA superfluid phases
	 from the Mott phase are indistinguishable on the large-scale
	 plot. The Mott phase is located in the extremely narrow region between
	 the superfluid phases.
	 Plot (b), therefore, shows the difference $\delta q_1$ between
	 the two boundaries (solid line)
	 and the difference $\delta q_2$ between the line $J=0$ and the boundary separating
	 the ferromagnetic superfluid phase from the Mott phase (dashed line).
        }
\label{qfp}
\end{figure}

The change of the sign of $J$ is closely related to the form of the dispersion relation
$E(k)$ for the lowest Bloch band. Using the definition of the Wannier spinors
the expression for $J$ can be rewritten in the form
$J = \frac{1}{\pi} \int_0^{k_L} E'(k) \sin(\pi k/k_L) \, dk$.
In the case of one-component condensate one always has a normal dispersion,
i.e., $E'(k)\ge 0$ for $0 \le k \le k_L$.
In the case we are dealing with one can get anomalous dispersion,
i.e., $E'(k) < 0$ for $0 \le k \le k_L$, as well. The change of the dispersion
type and as a consequence of the sign of $J$ happens at the points $(q,\theta)$
indicated in Figs.~\ref{qf} and~\ref{qfp}.
Since the type of dispersion is different in the ferromagnetic and
antiferromagnetic superfluid phases, one can expect principally different
properties of the nonlinear atomic matter waves in the two regimes~\cite{TS}.

The phase diagram of the Hamiltonian (\ref{BHH}) is determined by the ratio
$J/U$. In the case of one-component BECs it is a monotonic function of $|q|$.
In the two-component case we are dealing with, it has quite different properties.
Its dependence on $q$ and $\theta$ is not monotonic, and it can vanish or
even change its sign at certain finite values of $q$.
Therefore, it is reasonable to draw $\mu-q-\theta$ diagrams, which are shown in the figures,
instead of $\mu-J$ diagrams.
In the mean-field approximation and second-order perturbation theory
the boundary between the superfluid and the Mott regions is given
by~\cite{Sachdev}
$
2|J|/U
=
\left(
    1-n+\mu/U
\right)
\left(
    n-\mu/U
\right)
/
\left(
    1+\mu/U
\right)
$,
where $n-1<\mu/U<n$.
The critical values
$
\left(
    2|J|/U
\right)_c
=
1+2n-2\sqrt{n^2+n}
$,
which correspond to
$
\left(
    \mu/U
\right)_c
=
\sqrt{n^2+n}-1
$,
define the maximal possible width of the Mott region on the phase diagram.
Since $J/U$ is not a monotonic function of $q$ and $\theta$, there can be
several critical values of $q$ and $\theta$ (see
Figs.~\ref{phdq},~\ref{phdf}, and~\ref{phd-pq}).
At the points $(q,\theta)$, where $J/U$ vanishes, we have the Mott phase for any
values of $\mu/U$. These points define the boundary between the ferromagnetic
and antiferromagnetic states.

The phase diagrams in the plane spanned by $\theta$ and $q$
for $(\mu/U)_c=\sqrt{2}-1$ at $n=1$ and negative (red)
and positive (blue) detuning $\Delta$
are shown in Figs.~\ref{qf} and~\ref{qfp}, respectively.
The principal difference between the cases of positive and negative $\Delta$
can be understood if we apply the unitary transformation
\begin{equation}
\label{U}
\hat U
=
\frac{1}{\Omega}
\left(
    \begin{array}{rr}
        \Omega_+ & \Omega_- \\
       -\Omega_- & \Omega_+
    \end{array}
\right)
\,,
\Omega
=
\sqrt{\Omega_+^2 + \Omega_-^2}
\end{equation}
to the Wannier spinors in Eq.(\ref{BHH}).
After the tranformation we end up with the bright and dark states~\cite{DO},
which are not degenerate in contrast to the original ones.
The important point is that only the bright state is directly coupled to
the electromagnetic field and influenced by the potential
\begin{equation}
\label{VB}
V_B
=
\hbar\frac{\Omega^2}{\Delta}
=
\hbar\frac{\Omega_0^2}{\Delta}
\left(
    1+\cos\theta \cos 2 k_L z
\right)
\,.
\end{equation}

We consider first the case $\theta=0$, when the transformation $\hat U$
does not depend on the position $z$. In this case the dark state does not
``feel" any periodic potential. Since in the case $\Delta>0$ the dark state
has a lower energy the atoms stay in this state and, therefore, they can not
be localized on the lattice sites for any laser intensity.
In the case $\Delta<0$ the situation is reversed:
The energy of the bright state is lower than that of the dark one and only
the bright state is populated by the atoms. Therefore, increasing the laser intensity,
one can strongly localize the atoms on the lattice sites in exactly the same
manner as in the case of the one-component BEC.

In the case $\theta \neq 0$, $\hat U$ is a position-dependent transformation.
The atomic center-of-mass motion leads to the gauge potential
\begin{equation}
\label{Vg}
V_g
=
\hbar\omega_R
\left(
    \frac{\sin\theta}{1+\cos\theta \cos 2 k_L z}
\right)^2
\,,
\end{equation}
acting on the bright and dark atomic states and to the motional coupling
of the states~\cite{DO}. It seems that the transformation $\hat U$ does not lead to any
simplifications in the case $\theta \neq 0$. Nevertheless, it allows one to understand
what is going on, assuming that
$
\left|
    \Omega_0^2/\Delta
\right|
\gg
\omega_R
$.
In this approximation $V_g \ll \left|V_B\right|$,
and one can neglect the gauge potential for the bright state as well as
the motional coupling between the bright and dark
states~\cite{DO}. Then the only potentials acting on the bright and dark states are
given by Eqs.~(\ref{VB}) and ({\ref{Vg}}), respectively.
On the basis of the same argument as in the case $\theta=0$ we see that the atomic
localization in the cases $\Delta<0$ and $\Delta>0$ is determined by the potentials
$V_B$ and $V_g$, respectively.

Accordingly in the case $\Delta<0$ the quantity $\Omega_0^2 \cos(\theta)/\Delta$ plays the role
of the effective Rabi frequency and defines the strength of the periodic potential.
Therefore, in the simple approximation we consider for the sake of this
discussion, the critical effective Rabi frequency
$
\Omega_{eff}^c
=
\tilde\Omega_{eff}^c/2\cos\theta
$
(dotted line in Fig.~\ref{qf}), where $\tilde\Omega_{eff}^c$ is the critical
effective Rabi frequency of the one-component BEC and the factor 2 reflects
the fact that two atomic ground states are involved.
Since $\tilde\Omega_{eff}^c \sim 10\,\omega_R$~\cite{Jaksch},
the potential $V_g$ appears to be too weak and the transition into the Mott state
in the case $\Delta>0$ is not possible
(compare with the exact results in Figs.~\ref{phd-pq} and~\ref{qfp}).

\psfrag{g}[c]{\rotatebox{180}{$\delta q$}}

As it follows from the results presented in Figs.~\ref{qf} and~\ref{qfp},
this simplified description provides a correct physical insight,
but does not always work and at the laser intensities we are dealing with
one has to take into account the coupling between the components,
as we have done in the numerical calculations leading to the results displayed in
Figs.~\ref{phdq}--\ref{qfp}.
It is also necessary to note that
the one-component approximation cannot describe the change of sign in $J$ and
the related transition from the ferromagnetic to antiferromagnetic
superfluid phase.

Summarizing, we have studied quantum phase transitions of a BEC with two degenerate ground
states in an optical lattice created by the lin-angle-lin laser configuration.
It is shown that the periodic coupling of the atomic ground states modifies essentially
the phase diagram of the system. The most surprising feature we
encountered are
(i) change of sign of the tunneling matrix element $J$ under
variation of the laser intensity and the angle $\theta$, which leads to the ferromagnetic
and antiferromagnetic phase ordering analogous to the spin ordering
in magnetic systems, and
(ii) suppression of the Mott transition in the case $\Delta>0$,
except for a narrow region in the phase diagram (Figs.~\ref{phd-pq} and~\ref{qfp}).
Numerical results presented in the figures are
for the case of a commensurate filling of the lattice with one atom
per site. For higher occupation numbers the physics remains the same and there are
only minor quantitative changes.

\acknowledgments

This work has been supported by the SFB/TR 12
``Symmetries and universalities in mesoscopic physics".


\end{document}